\documentclass[12pt]{nostro}

\usepackage{epsfig}\parskip 5pt plus 1pt
\renewcommand{\>}{\rangle}
\newcommand{\be}{\begin{equation}}
\newcommand{\ee}{\end{equation}}
\newcommand{\bea}{\begin{eqnarray}}
\newcommand{\eea}{\end{eqnarray}}

\def\simge{\mathrel{%
   \rlap{\raise 0.511ex \hbox{$>$}}{\lower 0.511ex \hbox{$\sim$}}}}
\def\simle{\mathrel{
   \rlap{\raise 0.511ex \hbox{$<$}}{\lower 0.511ex \hbox{$\sim$}}}}

\begin{document}
\begin{frontmatter}
\thispagestyle{empty}
\begin{flushright}
{IFT-UAM/CSIC-06-15}
\end{flushright}
\title{Alternating ions in a $\beta$-beam to solve degeneracies} 
\author[Madrid]{A. Donini}
\author[Madrid]{E. Fern\'andez-Mart\'{\i}nez}
\address[Madrid]{I.F.T. and Dep. F\'{\i}sica Te\'{o}rica, Univ. Autonoma Madrid., E-28049, Madrid, Spain}
\vspace{.3cm}
\begin{abstract}
We study how the eightfold-degeneracy in the ($\theta_{13},\delta$) plane observed in $\gamma \sim 100$ $\beta$-beams 
can be reduced by periodically changing  the ions in the storage ring. This ``ions cocktail'' allows to change 
the neutrino energy, at fixed $\gamma$, by choosing ions with different decay energies. We propose to combine the 
standard $^6$He and $^{18}$Ne beams with $^8$Li and $^8$B ones. 
These latter two ions have peaked $\nu_e \to \nu_\mu$ oscillation probabilities for $\gamma = 100$ at a baseline 
$L \sim 700$ Km. At this distance and this $\gamma$ the oscillation probability of $^6$He and $^{18}$Ne neutrinos is at its second maximum. 
This setup is particularly suited for large enough values of $\theta_{13}$ (within reach at T2K-I) and it allows solving 
most of the eightfold-degeneracy, measuring $\theta_{13}, \delta$ and the sign of the atmospheric mass difference 
for values of $\theta_{13} \geq 5^\circ$. 
\end{abstract}

\vspace*{\stretch{2}}
\begin{flushleft}
\end{flushleft}
\end{frontmatter}

\section{Introduction}

The results of atmospheric, solar, accelerator and reactor neutrino experiments \cite{exp} show 
that flavour mixing occurs not only in the hadronic sector, as it has been known for long, but 
in the leptonic sector as well. 
The full understanding of the leptonic mixing matrix constitutes, together with the discrimination 
of the Dirac/Majorana character of neutrinos and with the measurement of their absolute mass scale, the main 
neutrino-physics goal for the next decade. 

The experimental results point to two very distinct mass 
differences\footnote{A third mass difference, $\Delta m^2_{LSND} \sim 1$ eV$^2$, suggested by 
the LSND experiment \cite{lsnd}, has not being confirmed yet \cite{boone} and will not be 
considered in this paper.}, $\Delta m^2_{sol} \approx 7.9 \times 10^{-5}$ eV$^2$ and 
$|\Delta m^2_{atm}| \approx 2.4 \times 10^{-3}$ eV$^2$.
On the other hand, only two out of the four parameters of the three-family leptonic mixing matrix $U_{PMNS}$ 
\cite{neutrino_osc} are known: $\theta_{12} \approx 34^\circ$ and $\theta_{23}\approx 41.5^\circ$  \cite{Fogli:2005cq}. 
The other two parameters, $\theta_{13}$ and $\delta$, are still unknown: for the mixing angle
$\theta_{13}$ direct searches at reactors \cite{chooz} and three-family global analysis of the 
experimental data give the upper bound $\theta_{13} 
\leq 11.5^\circ$, whereas for the leptonic CP-violating phase $\delta$ we have no information 
whatsoever. Two additional discrete unknowns are the sign of the atmospheric mass difference, 
$s_{atm}= $sign($\Delta m^2_{atm}$), and the $\theta_{23}$-octant, $s_{oct}= $sign($\tan 2 \theta_{23}$). 
The two unknown parameters $\theta_{13}$ and $\delta$ can be measured in ``appearance'' experiments 
through $\nu_e \to \nu_\mu, \nu_\mu \to \nu_e$ (the ``golden channel'', 
\cite{Cervera:2000kp}) and $\nu_e \to \nu_\tau$ (the ``silver channel'', \cite{Donini:2002rm}) oscillations. 
However, strong correlations between $\theta_{13}$ and $\delta$ and the presence of parametric degeneracies 
in the ($\theta_{13},\delta$) parameter space, \cite{Burguet-Castell:2001ez}-\cite{Barger:2001yr}, make 
the simultaneous measurement of the two variables extremely difficult\footnote{
A further problem is our present imprecise knowledge of atmospheric parameters, whose uncertainties
are far too large to be neglected when looking for the $\nu_\mu \to \nu_e$ and $\nu_e \to \nu_\mu, \nu_\tau$ 
oscillation probabilities \cite{Donini:2005rn}.}.
Most proposed solutions to these problems suggest the combination of different experiments and facilities, such as
Super-Beam's (of which T2K \cite{Itow:2001ee} is the first approved one), $\beta$-beam's \cite{Zucchelli:sa} 
or the Neutrino Factory \cite{Geer:1997iz}.

In this letter we propose to alleviate the parametric degeneracy in the ($\theta_{13},\delta$) plane by 
combining $\beta$-beam's obtained from the decay of several different ions. 

The $\beta$-beam concept was first introduced in Ref.~\cite{Zucchelli:sa} (see Ref.~\cite{Volpe:2006in} for a recent review).  
It involves producing a huge number of $\beta$-unstable ions, accelerating them to some reference
energy, and allowing them to decay in the straight section of a storage ring, resulting in a very intense and pure 
$\nu_e$ or $\bar \nu_e$ beam. ``Golden'' sub-leading transitions, $\nu_e \to \nu_\mu$ and $\bar \nu_e \to \bar \nu_\mu$,
can then be measured through muon observation in a distant detector.
The $\beta$-beam concept shares with the Neutrino Factory two main advantages with respect to conventional beams
(where neutrinos are obtained via pion decay): a) the neutrino flux is pure (for a $\beta$-beam, only $\nu_e$ or $\bar \nu_e$ 
neutrinos are present in the flux), thus decreasing the beam-induced background, 
and b) the neutrino spectrum can be exactly computed, thus strongly reducing flux systematics.

The neutrino flux per solid angle in a detector located at distance $L$ from the source, aligned with the boost direction 
of the parent ion is~\cite{Burguet-Castell:2003vv}:

\be
\left.{d\Phi \over dS dy}\right|_{\theta\simeq 0} \simeq
{N_\beta \over \pi L^2} {\gamma^2 \over g(y_e)} y^2 (1-y)
\sqrt{(1-y)^2 - y_e^2}, 
\label{eq:flux}
\ee
where $0 \leq y={E \over 2 \gamma E_0} \leq 1-y_e$, $y_e=m_e/E_0$
and
\be
g(y_e)\equiv {1\over 60} \left\{ \sqrt{1-y_e^2} (2-9 y_e^2 - 8
y_e^4) + 15 y_e^4 \log \left[{y_e \over
1-\sqrt{1-y_e^2}}\right]\right\} \, ,
\ee
where $\gamma$ is the ion boost factor.
In this formula $E_0$ represents the electron end-point energy, $m_e$ the electron mass, $E$ the energy of the final state 
neutrino and $N_\beta$ the total number of ion decays per year.
An important difference of the $\beta$-beam flux with respect to the Neutrino Factory is that only $\nu_e (\bar \nu_e)$ are 
present in the beam, and thus final lepton charge identification is not needed. This permits to use large 
water \v Cerenkov detectors at a $\beta$-beam, something impossible at the Neutrino Factory, where magnetized detectors 
are mandatory. 

\section{The Alternating Ions Scheme}

The key parameter in the optimization of the $\beta$-beam flux is the relativistic $\gamma$ factor: if the 
baseline is tuned to be at an oscillation peak for $\nu_e \to \nu_\mu$ transitions, indeed, the statistics
that can be collected in the detector scales linearly with $\gamma$ \cite{Zucchelli:sa}.
This can be derived from eq.~(\ref{eq:flux}) as follows: in the hypothesis of linear dependence of the neutrino-nucleon 
cross-section on the neutrino energy and for $L/E$ tuned to the $n$-th $\nu_e \to \nu_\mu$ oscillation peak, 
the number of events expected in the far detector located at distance $L$ is: 
\be
N_{events} \propto N_\beta \left ( \frac{\Delta m^2}{2 n -1} \right )^2 \frac{\gamma}{E_{cms}}
\label{eq:stats}
\ee
where $E_{cms}$ is the mean neutrino energy in 
the center-of-mass system of the $\beta$-decay (with $\<E\> = 2 \gamma E_{cms}$). 
Applying this formula, the signal statistics in the far detector increases linearly with 
the boost factor $\gamma$ and the number of decays per year $N_\beta$, and decreases linearly with 
the mean neutrino energy. 

Three key features must be considered when choosing the optimal $\beta$-emitters: 
a) from eq.~(\ref{eq:stats}), it can be seen that the lower the mean neutrino energy, the larger 
the statistics\footnote{Notice, however, that this formula is not appropriate for neutrino energies below 1 GeV, 
where the cross-section energy dependence is $E^k$ with $k \ge 1$. 
This is, on the other hand, the typical range of energies considered for ``low'' $\gamma$ $\beta$-beams.};
b) assuming a limited space charge capacity of the storage ring, low-Z isotopes can be stored in larger number
than high-Z isotopes \cite{Autin:2002ms}; 
c) the ion half-life $T_{1/2}$ must be long enough to accelerate the ions to the desired energy and short enough 
to allow a large number of them to decay in the storage ring such as to produce an intense neutrino beam. 

In Tab.~\ref{tab:ions} we remind the relevant parameters for five ions: $^{18}$Ne, $^{19}$Ne and $^6$He, $^8$Li and $^8$B.
\begin{table}
\begin{center}
\begin{tabular}{|c|c|c|c|c|} \hline \hline
   Element  & $A/Z$ & $T_{1/2}$ (s) & $Q_\beta$ eff (MeV) & Decay Fraction \\ 
\hline
  $^{18}$Ne &   1.8 &     1.67      &        3.41         &      92.1\%    \\
            &       &               &        2.37         &       7.7\%    \\
            &       &               &        1.71         &       0.2\%    \\ 
  $^{19}$Ne &   1.9 &    17.2       &        2.31         &       100\%    \\
  $^{8}$B   &   1.6 &     0.77      &       13.92         &       100\%    \\
\hline
 $^{6}$He   &   3.0 &     0.81      &        3.51         &       100\%    \\ 
 $^{8}$Li   &   2.7 &     0.83      &       12.96         &       100\%    \\ 
\hline
\hline
\end{tabular}
\caption{\it $A/Z$, half-life and end-point energies for three $\beta^+$-emitters ($^{18}$Ne, $^{19}$Ne and $^8$B)
and two $\beta^-$-emitters ($^6$He and $^8$Li). All different $\beta$-decay channels for $^{18}$Ne are presented~\cite{betadecays}.}
\end{center}
\label{tab:ions} 
\end{table}
Consider first $^6$He and $^{18}$Ne: as it was stressed in the literature (starting with Ref.~\cite{Zucchelli:sa}), $^6$He 
has the right half-life to be accelerated and stored such as to produce an intense $\bar \nu_e$ beam. 
This is also the case for $^{18}$Ne, that has been shown to be the best candidate as $\beta^+$-emitter. 
Other ions such as $^8$B and $^{33}$Ar were discarded for different reasons: the former is difficult to produce with 
standard ISOLDE techniques (it reacts with many elements typically used in ISOL targets and ion sources and is 
therefore barely released); the latter is too short-lived to be accelerated to the desired energy ($T_{1/2} = 0.17$ s). 

The goal luminosity needed for physics have been fixed to $2.9 \times 10^{18}$ ions-decay per year for the $\beta^-$-emitter and 
$1.1 \times 10^{18}$ ion-decays per year for the $\beta^+$-emitter \cite{Bouchez:2003fy}. 
The EURISOL collaboration has recently reported the following preliminary results from the EURISOL DS~\cite{Lindroos}, 
using a 2.2 GeV proton driver with a 0.10 mA current and 220 kW power (from the SPL Design Report, \cite{Autin:2000mn}):
\begin{itemize}
\item {\bf $^6$He} \\
Using an ISOL BeO target, $5 \times 10^{13}$ atoms per second are produced and the goal $\bar \nu_e$ luminosity is attained.
The $^6$Li ions produced in the $\beta^-$-decay of $^6$He could in principle interact with 
the storage ring magnets, thus producing an undesired $\nu_\mu$ and $\bar \nu_\mu$ beam-background. 
This background was studied in Ref.~\cite{Zucchelli:sa} and it is at the level of $10^{-4}$.
\item {\bf $^{18}$Ne} \\
Using an ISOL MgO target, $2 \times 10^{12}$ atoms per second are produced. It is supposed that the goal luminosity for 
the $\nu_e$ beam is also attainable. $^{18}$Ne undergoes $\beta^+$-decay into three different $^{18}$F states (as reported in Tab.~\ref{tab:ions}).
The beam background for this ion has not been studied in detail, but it is considered to be similar to that of $^6$He, and 
thus negligible.
\item {\bf $^{19}$Ne} \\
This ion has been studied as an alternative $\beta^+$-emitter. It has $A/Z$ and end-point energy similar to $^{18}$Ne but 
a far longer half-life. In this case $4 \times 10^{13}$ atoms per second are supposed to be produced and, thus, reaching 
the goal luminosity should be at hand. Also, the target production performance for this ion is higher than for $^{18}$Ne: 
12\% versus 4\% (to be compared, however, with the $^6$He target production performance of 100\%), \cite{Lindroos}. 
\end{itemize}

The two ions that we propose as an alternative to $^6$He and $^{18}$Ne (or $^{19}$Ne) as $\beta^-$- and $\beta^+$-emitters are 
$^8$Li and $^8$B, respectively. $^8$Li has similar half-life, $Z$ and $A/Z$ to $^6$He, thus sharing the key characteristics needed 
for the bunch manipulation. $^8$B has a shorter lifetime than $^{18}$Ne, similar $A/Z$ and $Z$ much smaller than $^{18}$Ne 
(which could in principle allow to store a larger amount of ions in the storage ring).
Both ions have a much larger end-point energy than the two reference ions or $^{19}$Ne.
As a consequence, for a fixed $\gamma$, a longer baseline is needed to tune the $L/E$ ratio to the first oscillation
peak with respect to $^6$He or $^{18}$Ne beams, and thus a smaller signal statistics is expected in the far detector. 
Therefore, the expected sensitivity to $\theta_{13}$ of such beams is smaller than that for a ``standard'' beam 
with $\bar \nu_e$ and $\nu_e$ produced via $^6$He and $^{18}$Ne with a baseline tuned to the first oscillation peak. 
Notice, however, that being the ion-decay end-point energy for $^8$Li and $^8$B larger, a smaller $\gamma$ is needed to reach the 
desired boosted neutrino energy. 
Since the $\gamma$ choice depends in last instance on the facility that is used to accelerate the ions, it is then possible to reach
higher neutrino energies using the same facility to accelerate the ions to be stored (see Ref.~\cite{Donini:Hamburg}). 
If we combine different ions, we can (using the same facility and the same $\gamma$ factor) produce neutrino beams of 
different $L/E$ that can be used to disentangle many of the parametric degeneracies discussed before. 
As it was shown in the literature (see, for example, Refs.~\cite{Donini:2003vz}-\cite{Donini:2004iv})
degeneracies are indeed best lifted combining beams with different $L/E$. 

When we first proposed to produce neutrino beams from different ions to soften the parametric degeneracy in ($\theta_{13},\delta$) 
\cite{Donini:Hamburg}, $^8$Li came up immediately as a good $\beta^-$-emitter to be combined with $^6$He \cite{Autin:2002ms}. 
However, no reasonable $\beta^+$-emitter to be combined with $^{18}$Ne was identified at that time.
We considered the idea not mature enough to be presented as an alternative to increasing $\gamma$ with fixed ion composition, 
such as in Ref.~\cite{Burguet-Castell:2003vv}. Things changed with Ref.~\cite{Rubbia:2006pi}, where it was shown 
that intense $^8$B (and $^8$Li) fluxes can be achieved with the {\it ionisation cooling} technique. 

\begin{itemize}
\item {\bf $^8$Li} \\
It is not difficult to produce an intense $\bar \nu_e$ beam from $^8$Li decay, since with standard ISOL targets
the ion production rate is much larger for $^8$Li than for $^6$He: we have $6 \times 10^8$ ions per $\mu$C in the case of $^8$Li to be compared
with $6 \times 10^6$ ions per $\mu$C in the case of $^6$He \cite{Tengblad}. 
The ISOL target used to produce $^8$Li is a thin Ta foil. 
The $\beta^-$-decay channel of $^8$Li is $^8$Li $\to$ $^8$Be $\to$ 2$\alpha$. The two $\alpha$'s are then immediately bended 
by the storage ring magnetic field (being much lighter than the circulating ions) and no beam-background is expected.
We will consider in the rest of the paper a $\bar \nu_e$ flux from $^8$Li decay of $2.9 \times 10^{18}$ ion-decays per year,
comparable with that attainable with $^6$He decay.
\item {\bf $^8$B} \\
The case of $^8$B is different: this ion was previously discarded as a $\beta^+$-emitter since it is extremely difficult 
to produce at a sufficient rate with ISOLDE techniques. However, using the {\it ionisation cooling} technique \cite{Rubbia:2006pi},
sustained $^8$Li and $^8$B production is supposed to be at reach through the reactions $^7$Li + D $\to$ $^8$Li + p and 
$^6$Li + $^3$He $\to$ $^8$B + n. $^8$B also decays into $^8$Be and finally to two $\alpha$'s (it is the mirror of $^8$Li), 
and also in this case no beam-background is expected. It must be reminded, however, that the $^8$B $\beta$-decay spectrum
is affected by several systematics errors that must be tamed before using it for a precision experiment (see Ref.~\cite{Winter:2003ac}).
We will consider in the rest of the paper a $\nu_e$ flux from $^8$B decay of $1.1 \times 10^{18}$ ions-decay per year,
comparable with that attainable with $^{18}$Ne decay.
\end{itemize}

Four classes of setups have been considered so far: $\gamma \simeq 10$ (``very low'' $\gamma$) \cite{Volpe:2003fi}, 
$\gamma \sim 100$ (``low'' $\gamma$), 
with a typical baseline of $O (100)$ Km \cite{Burguet-Castell:2003vv}, \cite{Bouchez:2003fy}, \cite{Donini:2004hu} and \cite{Donini:2004iv}; 
$\gamma \sim 300$ (``medium'' $\gamma$), with $L = O (700)$ Km \cite{Burguet-Castell:2003vv}, \cite{Burguet-Castell:2005pa}-\cite{Donini:2005qg}; 
and $\gamma \geq 1000$ (``high'' $\gamma$), with baselines of several thousands kilometers, 
comparable with those suggested for the Neutrino Factory, \cite{Burguet-Castell:2003vv},\cite{Huber:2005jk} 
and \cite{Terranova:2004hu}. 
The three $\gamma$ ranges are related to different CERN-based facilities: the SPS, with an ultimate $\gamma \leq 250$);
a refurbished SPS (a facility that could be needed for LHC upgrades), with $\gamma \leq 500$; and the LHC, 
with $\gamma \geq 1000$. This last option, however, has not been studied in detail and we will not consider it in the
following. 

The physics case in which we are interested is that of a relatively large $\theta_{13}$ that could be at reach at the T2K-phase I
experiment starting in 2009. At this experiment, a $\nu_\mu \to \nu_e$ signal can be observed in the Super-Kamiokande detector
if $\theta_{13} \geq 3^\circ$. In case of a positive signal, new experiments will be launched aiming at a precision measurement of $\theta_{13}$
and at the measurement of the leptonic CP violating phase $\delta$ and of the sign of the atmospheric
mass difference $\Delta m^2_{atm}$ (something out of reach of the T2K-I experiment, for which only a $\nu_\mu$ flux will 
be produced and that has too short a baseline, $L = 295$ Km, to take advantage of matter effects).
In this scenario, we will show that the {\it Alternating Ions Scheme} represents an interesting alternative to the standard
$\beta$-beams considered up to now (where only one type of ion is used to produce the $\nu_e$ or the $\bar \nu_e$ fluxes), if the 
$\gamma$ boost factor is limited to $\gamma \leq 250$ for technical reasons.

We will compare results obtained with three different setups: 
\begin{itemize}
\item {\bf Standard ``low'' $\gamma$ scenario} \\
$L = 130$ Km (CERN to Fr\'ejus); $\gamma_{^6 He} = \gamma_{^{18} Ne} = 120$. 
Both fluxes are tuned to be at the first oscillation peak. 
A given ion is accummulated in the storage ring for ten years.
\item {\bf Alternating ions ``low'' $\gamma$ scenario} \\
$L = 650$ Km (CERN to Canfranc); 
$\gamma_{^8 Li} = \gamma_{^8 B} = 100$; $\gamma_{^6 He} = \gamma_{^{18} Ne} = 120$.  
The $^8$Li and $^8$B fluxes are tuned at the first oscillation peak, whereas
the $^6$He and $^{18}$Ne fluxes are tuned at the second oscillation peak.
A given ion is accummulated in the storage ring for five years.
\item {\bf Standard ``medium'' $\gamma$ scenario} \\
$L = 650$ Km (CERN to Canfranc); $\gamma_{^6 He} = \gamma_{^{18} Ne} = 350$. 
Both fluxes are tuned to be at the first oscillation peak.
A given ion is accummulated in the storage ring for ten years.
\end{itemize}

In all scenarios, a $\bar \nu_e$ flux of $2.9 \times 10^{18} \bar \nu_e$ per year or 
a $\nu_e$ flux of $1.1 \times 10^{18} \nu_e$ per year is aimed at the distant detector, regardlessly of the decaying ion.
For a 5 year running time per each ion stored, the total luminosity is $1.45 \times 10^{19}$ $\bar \nu_e$ 
and $5.5 \times 10^{18}$ $\nu_e$ aimed at the far detector. 

Throughout this letter, we will consider a 1 Mton mass water \v Cerenkov detector (500 Kton fiducial mass),
with a UNO~\cite{Jung:1999jq} or MEMPHYS design~\cite{Campagne:2006yx}. At the considered neutrino energies, this detector is believed to show 
a rather good neutrino energy reconstruction capability, allowing for a significant background rejection. 
Migration matrices for this detector at the considered $\gamma$ values are taken\footnote{The mean energy of neutrino and 
antineutrino beams obtained from $^8$Li and $^8$B decays with a boost factor $\gamma = 100$ is indeed comparable with 
that of beams obtained from $^6$He and $^{18}$Ne decays with a boost factor $\gamma = 350$, for which migration matrices 
are presented in that paper.} from Ref.~\cite{Burguet-Castell:2005pa}. 
Notice that this kind of detector could well be a non-optimal choice for the $L \sim 700$ Km baseline. If the $\beta$-beam
is located at CERN, this baseline roughly corresponds to the CERN-Gran Sasso or to the CERN-Canfranc distance.
For the former lab, it is certainly difficult to imagine a 1 Mton water \v Cerenkov detector sitting in one of the halls 
with no significant engineering effort. In that case a different detector can be the optimal choice 
(see Ref.~\cite{Donini:2005qg} for a medium $\gamma$ $\beta$-beam with a MINOS-like magnetized iron detector). 
However, we believe that to compare the standard ions choice with our Alternating Ions Scheme it is better 
to fix a detector technology in order to make an easy comparison between the two scenarios. 
In a forthcoming letter we will study an alternating ions setup better suited to a specific laboratory characteristics. 

\begin{figure}[t!]
\vspace{-0.5cm}
\begin{center}
\begin{tabular}{cc}
\hspace{-0.55cm} \epsfxsize7.5cm\epsffile{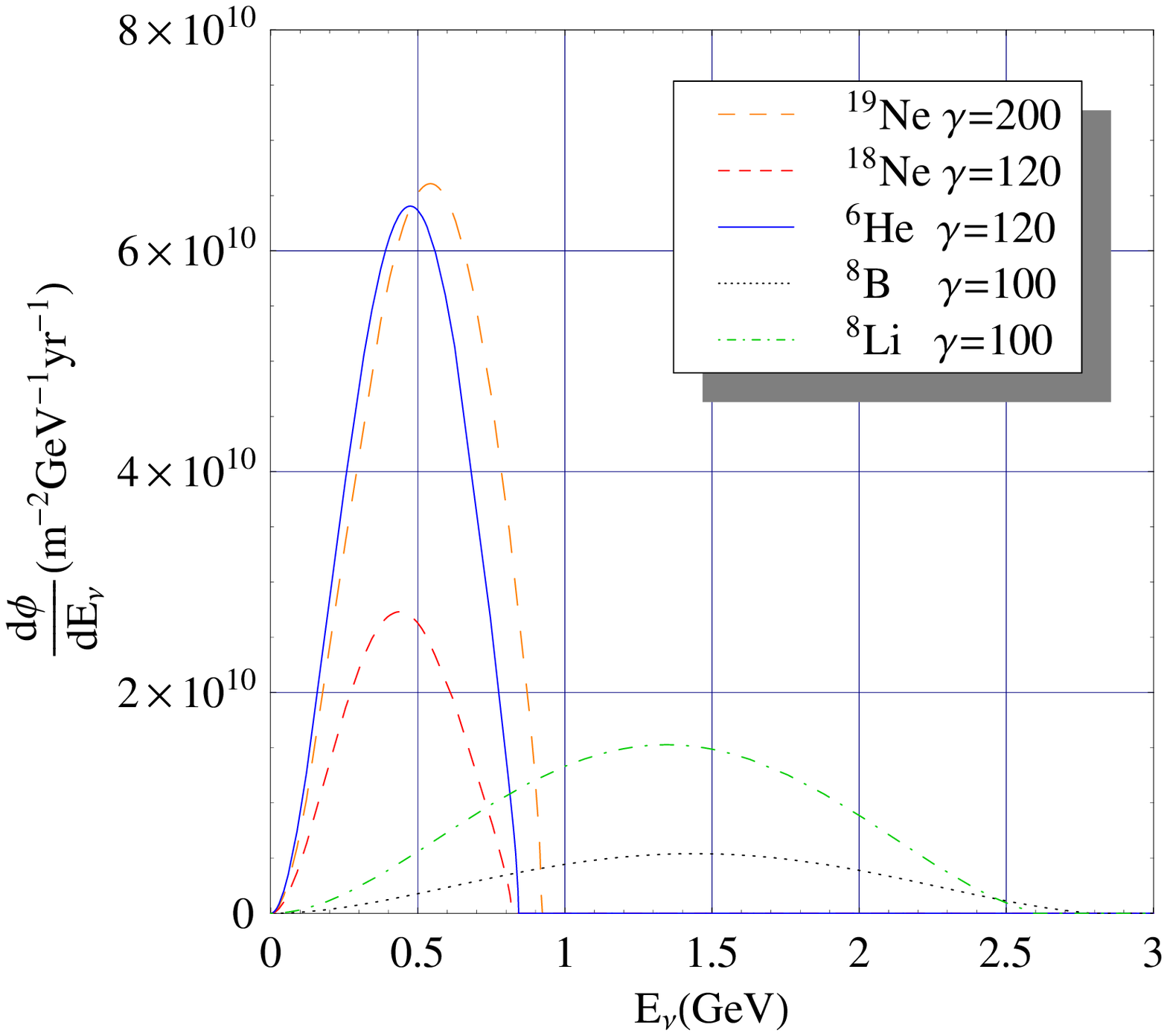} & 
                 \epsfxsize7.5cm\epsffile{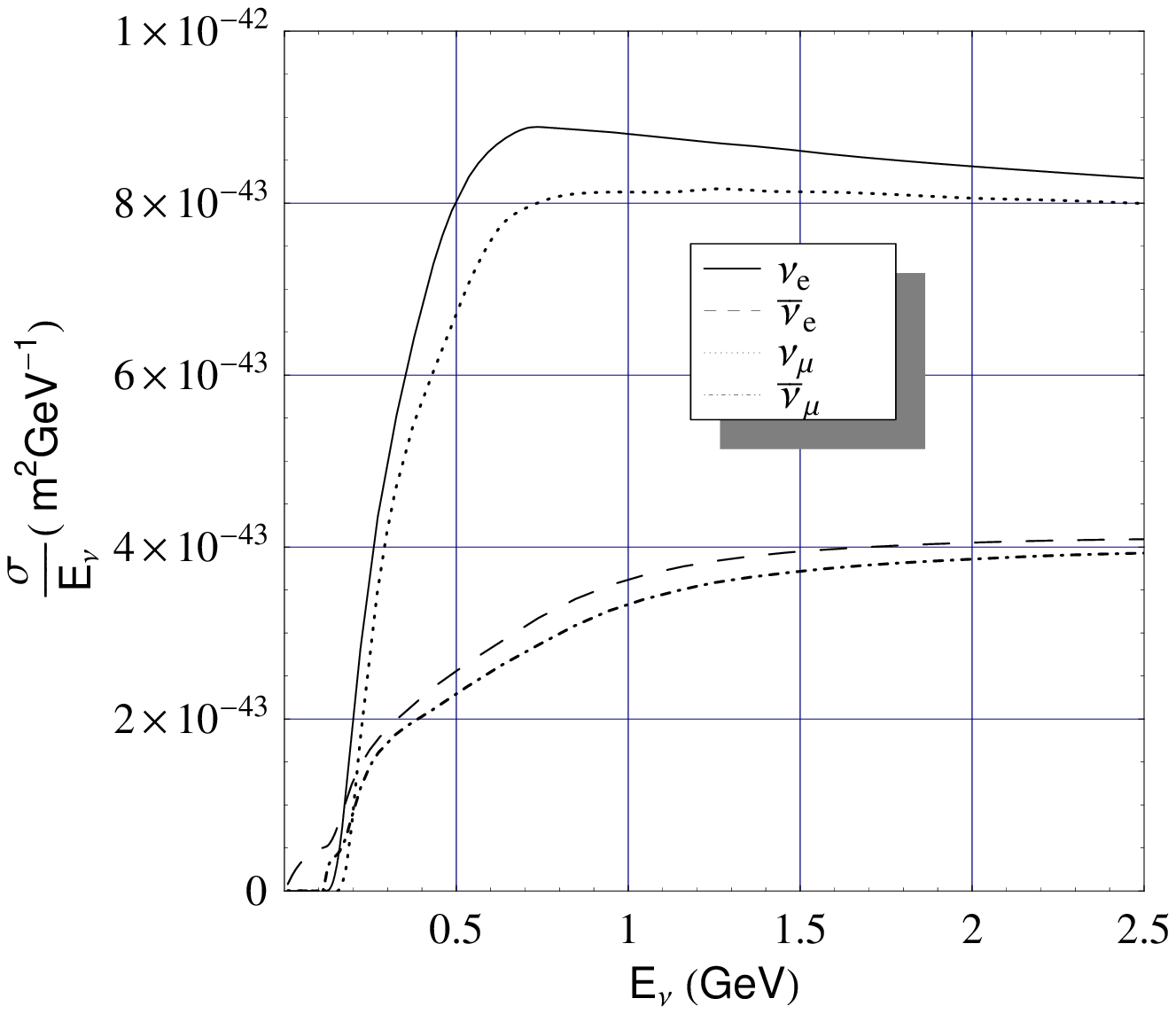}
\end{tabular}
\caption{\it 
Left: $\beta$-beam fluxes at a $L = 650$ Km baseline as a function of
the neutrino energy for $^{19}$Ne, $^{18}$Ne, $^6$He, $^8$B and $^8$Li. 
Right: the $ \nu N$ and $\bar \nu N$ cross-sections on water \cite{lipari}.}
\end{center}
\label{fig:fluxes}
\end{figure}

Fig.~\ref{fig:fluxes}(left) shows the $\beta$-beam neutrino fluxes computed at $L = 650$ Km from the source, 
keeping $m_e \neq 0$ and taking into account that $^{18}$Ne has different decay channels available. Be aware that even if $m_e$ effects seem
negligible, their inclusion could be sizable due to the dramatic cross-section suppression of low energy neutrinos.
In Fig.~\ref{fig:fluxes}(right) the $\nu$ and $\bar \nu$ cross-sections on water used in this letter, taken from Ref.~\cite{lipari}, are shown. 
Notice the difference between the $\nu_e N$ and $\bar \nu_e N$ cross-sections: 
the former, being an interaction between the $\nu_e$ and a neutron inside the oxygen nucleus, is affected by 
nuclear effects and thus shows a threshold energy. The latter is mainly a $\bar \nu_e$ interaction with the
protons of the two hydrogens, approximately free. This feature is quite relevant for neutrino/antineutrino of hundreds of MeV energy, 
region where different cross-sections can easily differ by a factor 2. Be aware that there are other nuclear effects
(see Ref.~\cite{Serreau:2004kx} and references therein) not included yet in any of the available calculations that could 
play an important effect at the cross-section threshold energy.

\begin{table}[hbtp]
\begin{center}
\begin{tabular}{|c|c|c|c|c|c|c|} 
\hline
$E$ (GeV)  & $^{18}$Ne$_{no-osc}$ & $^{18}$Ne$_{osc}$ & $^{18}$Ne$_{bkg}$ & $^6$He$_{no-osc}$ & $^{6}$He$_{osc}$ & $^6$He$_{bkg}$ \\ 
\hline
  0 - 0.5  &     2252             &      23           &     10            &       1937        &   36    &    12     \\
0.5 - 0.75 &     3309             &       9           &      3            &       3227        &   19    &     0     \\
0.75 - 0.9 &      318             &       1           &      0            &        527        &    1    &     0     \\
\hline
$E$ (GeV)  & $^8$B$_{no-osc}$ & $^8$B$_{osc}$ & $^8$B$_{bkg}$ & $^8$Li$_{no-osc}$ & $^8$Li$_{osc}$ & $^8$Li$_{bkg}$ \\ 
\hline
 0.5 - 0.75 &      439             &       4           &     18            &        546        &    6    &    23     \\
0.75 - 1.0  &      975             &       7           &      8            &       1341        &    7    &     4     \\
1.0 - 1.25  &     1604             &      10           &      4            &       2287        &   13    &     1     \\
1.25 - 1.50 &     2200             &      11           &      2            &       3060        &   16    &     1     \\
1.50 - 1.75 &     2613             &       8           &      0            &       3379        &   12    &     0     \\
1.75 - 2.0  &     2722             &       5           &      1            &       3074        &    9    &     0     \\
2.0 - 2.45  &     4037             &       3           &      0            &       3013        &    5    &     0     \\
\hline
\end{tabular}
\end{center}
\caption{\it 
Unoscillated CC events, signal and background rates per bin in the alternating ions ``low'' $\gamma$ setup
in a 500 Kton water \v Cerenkov detector located at $L = 650$ Km from the source, 
for 5 years of running time for each ion mode. The input parameters are $\theta_{13} = 5^\circ$ and $\delta = 0^\circ$.}
\label{tab:events}
\end{table}

In Tab.~\ref{tab:events} we present the expected signal and background in a 500 Kton fiducial mass water \v Cerenkov detector
located at $L = 650$ Km from the source, for 5 years of running time for each ion circulating in the storage ring.
The expected unoscillated CC electrons per bin are also reported. Efficiency is taken from Ref.~\cite{Burguet-Castell:2005pa}.
The oscillation parameters are \cite{Fogli:2005cq}: 
$\Delta m_{12}^2 = 7.9 \times 10^{-5}$ eV$^2$, $\theta_{12} = 34^\circ$; $|\Delta m_{23}^2| = 2.4 \times 10^{-3}$ eV$^2$, 
$\theta_{23} = 41.5^\circ$; 
$\theta_{13} = 5^\circ, \delta = 0^\circ$.  The sign of $\Delta m_{23}^2$ has been chosen to be positive.
Depending on the mean neutrino energy for the different beams, the signal is binned differently. 
For $^6$He and $^{18}$Ne (or $^{19}$Ne) beams (with mean energies $\<E_\nu \>= 0.44$ GeV and $\<E_{\bar \nu}\> = 0.46$ GeV), 
only three reconstructed energy bins are considered \cite{Burguet-Castell:2005pa}. On the other hand, for $^8$Li and $^8$B beams 
(with much larger mean energies, $\<E_\nu\> = 1.44$ GeV and $\<E_{\bar \nu}\> =  1.34$ GeV), 
up to seven reconstructed energy bins are considered \cite{Burguet-Castell:2005pa}.
We have included an overall 5\% gaussian systematic uncertainty. 

\section{Solving degeneracies}

\begin{figure}[t!]
\vspace{-0.5cm}
\begin{center}
\begin{tabular}{cc} 
\hspace{-0.55cm} \epsfxsize7.5cm\epsffile{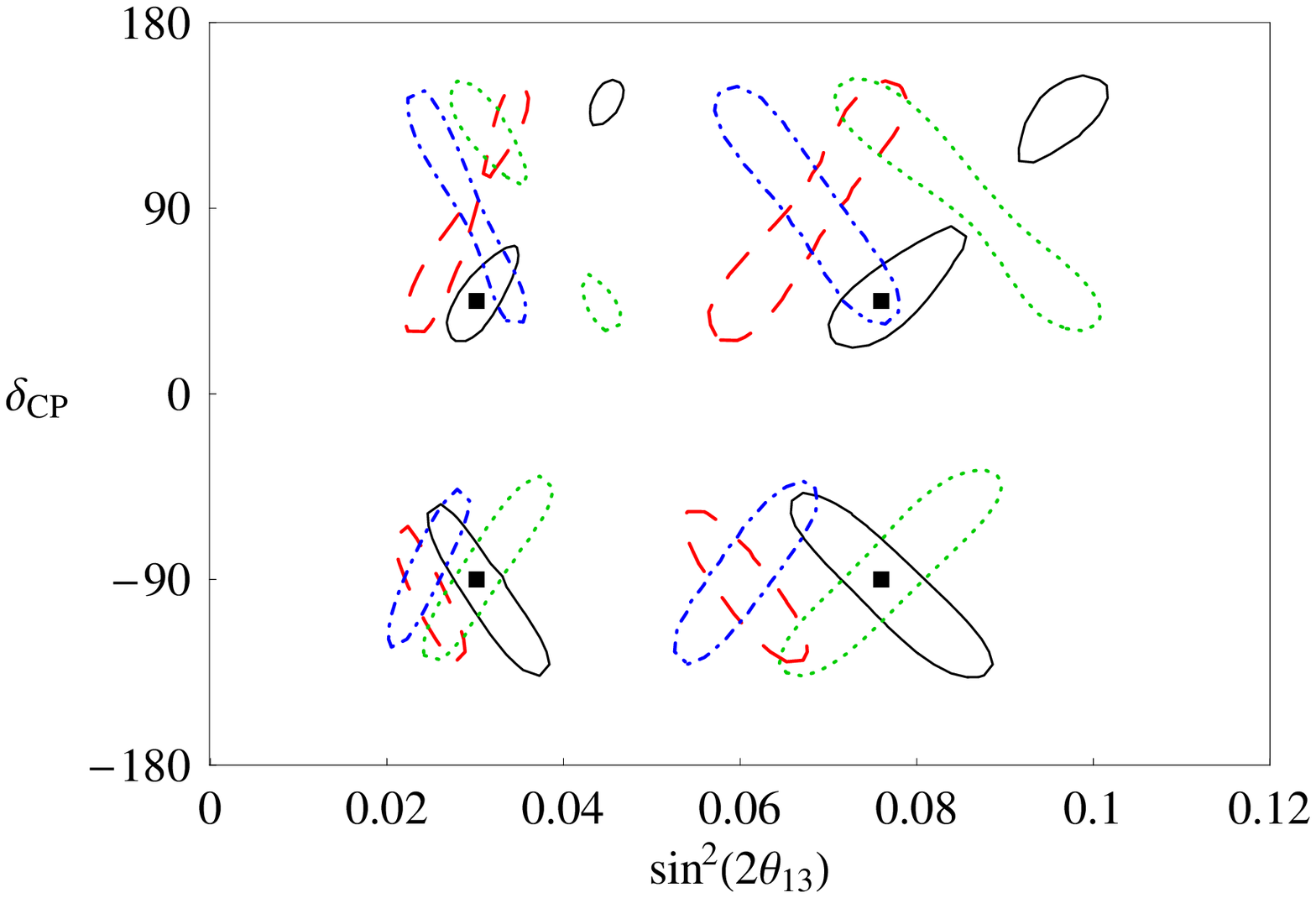} & 
                 \epsfxsize7.5cm\epsffile{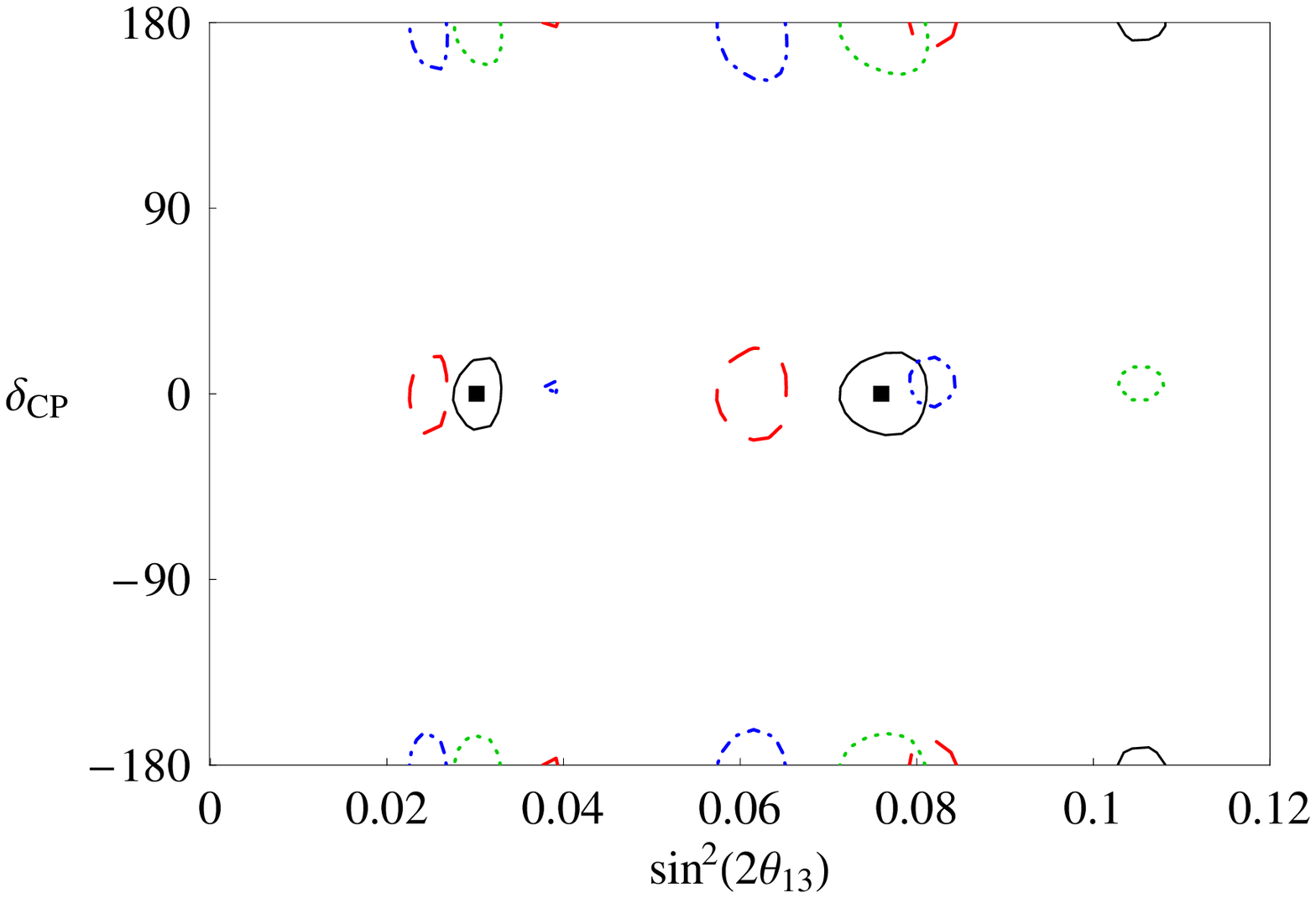} \\
\hspace{-0.55cm} \epsfxsize7.5cm\epsffile{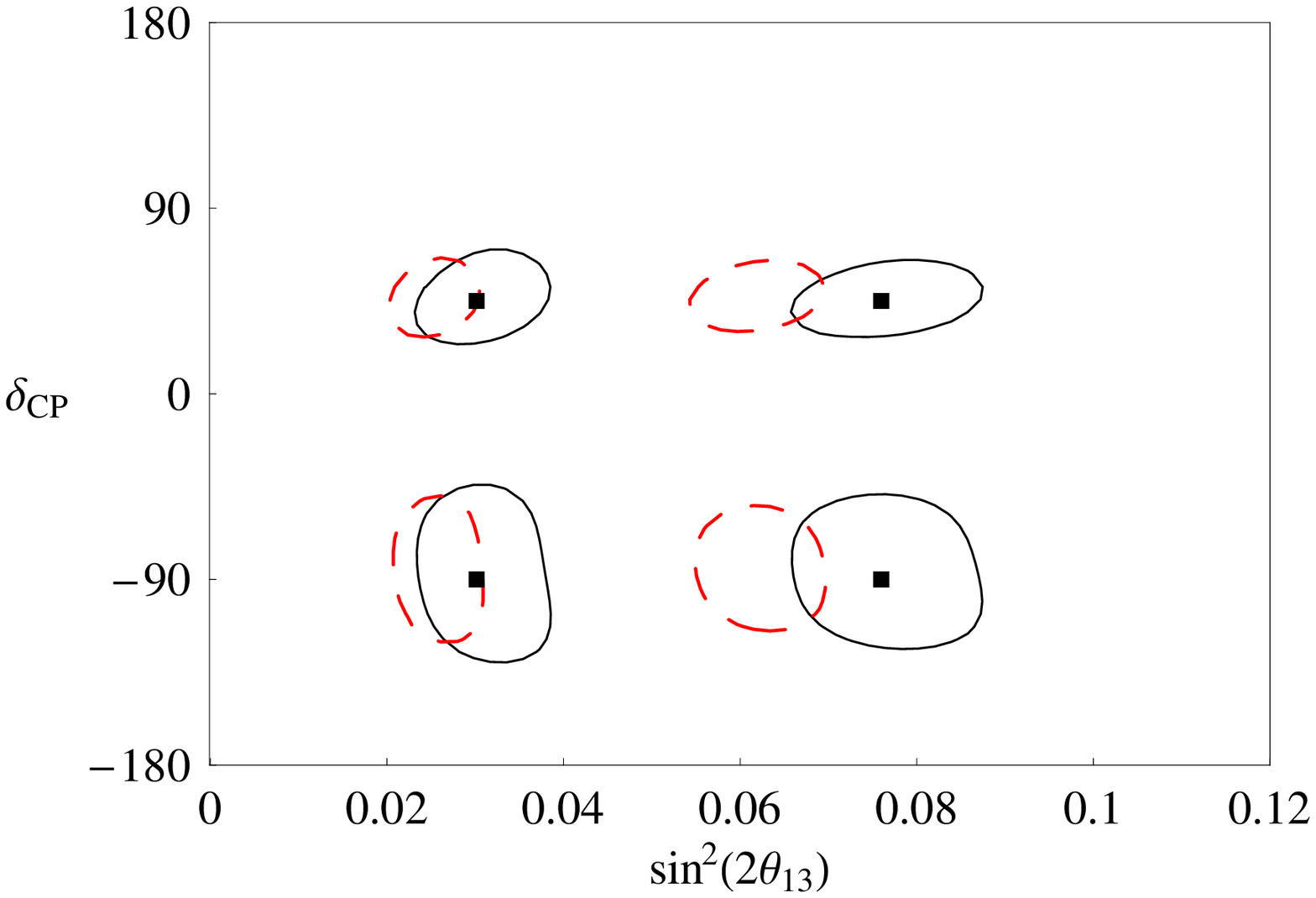} & 
                 \epsfxsize7.5cm\epsffile{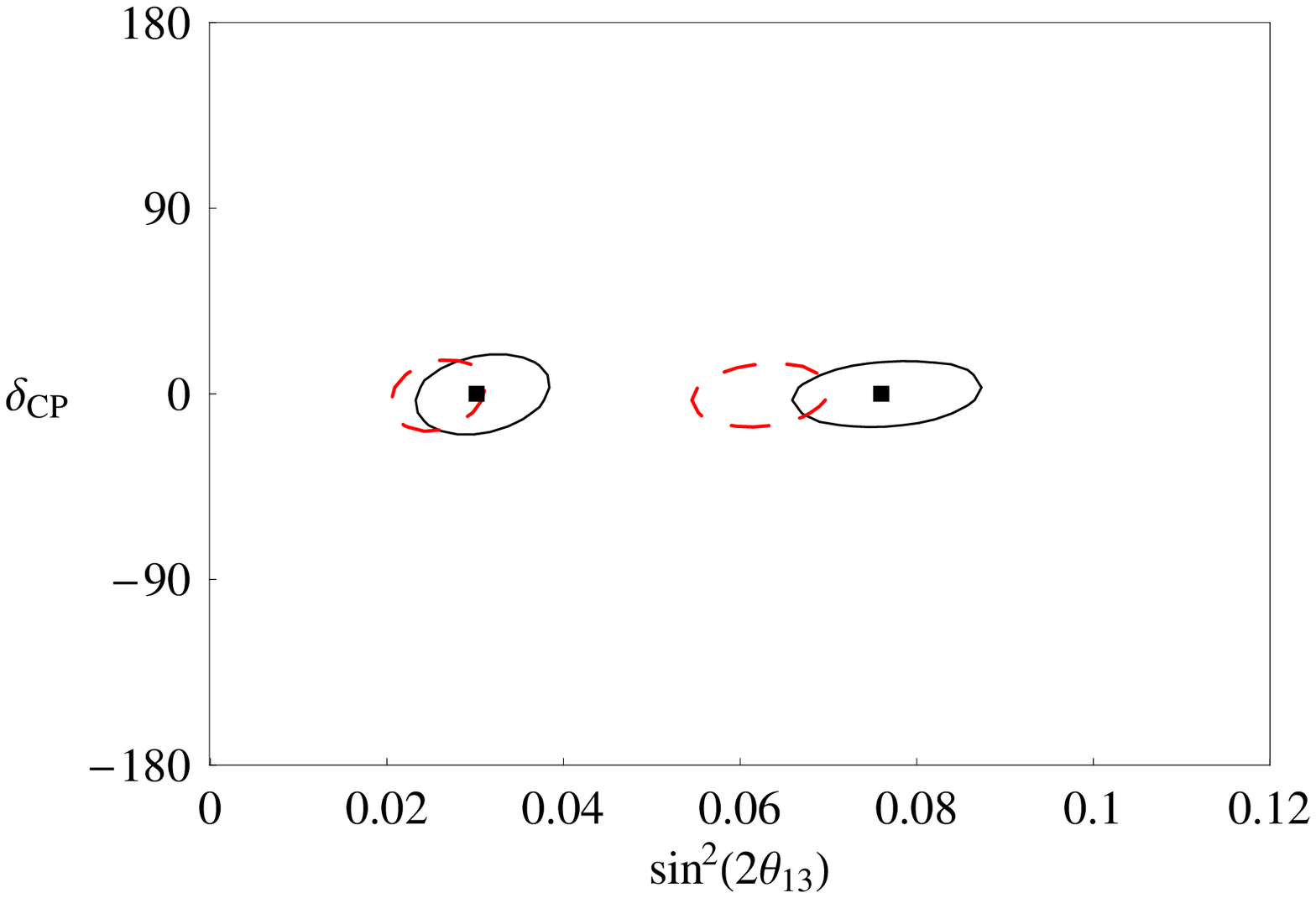}
\end{tabular}
\caption{\it 
90\% C.L. contours for the standard ``low'' $\gamma$ setup (top panels)
and the alternating ions ``low'' $\gamma$ setup (bottom panels): solid stands for the true solution and the intrinsic degeneracy; 
dotted for the ``sign clones''; dashed for the ``octant clones''; and dot-dashed for the ``mixed clones''. 
The input pair is labelled by a thick black square
and corresponds to $\delta = 45^\circ, -90^\circ$ (left panels) and $0^\circ$ (right panels), 
with $\theta_{13} = 5^\circ, 8^\circ$.
}
\label{fig:plots}
\end{center}
\end{figure}

In Fig.~\ref{fig:plots} we present our results for a fit to ``experimental data'' (generated as 
in \cite{Cervera:2000kp}) for the {\bf standard ``low'' $\gamma$ setup} (top panels)
and the {\bf alternating ions ``low'' $\gamma$ setup} (bottom panels). 
Different input values of $\delta$ are considered: $\delta = 45^\circ, - 90^\circ$ (left panels) and 
$\delta = 0^\circ$ (right panels). In each panel, two input values for $\theta_{13}$ (accessible at T2K-phase I)
are considered: $\theta_{13} = 5^\circ, 8^\circ$. The input pair is labelled by a thick black square. 
Lines represent 90\% C.L. contours: solid stands for the true solution and the intrinsic degeneracy; dotted stands for the ``sign clones''; 
dashed stands for the ``octant clones''; and dot-dashed stands for the ``mixed clones''.
The oscillation parameters are \cite{Fogli:2005cq}: 
$\Delta m_{12}^2 = 7.9 \times 10^{-5}$ eV$^2$, $\theta_{12} = 34^\circ$; $|\Delta m_{23}^2| = 2.4 \times 10^{-3}$ eV$^2$, 
$\theta_{23} = 41.5^\circ$.  The sign of $\Delta m_{23}^2$ has been chosen to be positive.
Since the sign of $\Delta m^2_{23}$ and the $\theta_{23}$-octant are unknown, fits to both sign[$\Delta m^2_{23}$] = $\pm$ 1 and
sign[$\tan (2 \theta_{23})$] = $\pm$ 1 have been performed (see Ref.~\cite{Donini:2004hu} for a description of parametric degeneracies at the
``low'' $\gamma$ $\beta$-beam). 

For both scenarios we have checked that employing $^{18}$Ne with $\gamma = 120$ or $^{19}$Ne with $\gamma = 200$ gives 
similar results (albeit slightly better in the latter case).

As it can be seen, for every considered input pair the {\it Alternating Ions Scheme} reduces the eightfold-degeneracy in the 
($\theta_{13},\delta$) plane to a twofold-degeneracy. The so-called ``intrinsic'' degeneracy \cite{Burguet-Castell:2001ez} 
is always solved (as it is usual when combining information from neutrino beams with different $L/E$). 
Most importantly, the sign of the atmospheric mass difference is measured. 
This is not possible at the standard ``low'' $\gamma$ setup since the baseline is too short 
to take advantage of matter effects to discriminate between hierarchical and inverted spectra. However, having a 
longer baseline is not enough: to solve the ``sign'' \cite{Minakata:2001qm} and the ``mixed'' \cite{Barger:2001yr}
degeneracies, thus measuring $s_{atm}$, it has been crucial to combine neutrino beams with different $L/E$ 
whose corresponding ``sign clones'' appear in different regions of the ($\theta_{13},\delta$) plane \cite{Enrique:Moriond}. 
The only surviving ambiguity is the ``octant'' degeneracy \cite{Fogli:1996pv}. 
Notice, however, that our ignorance on the $\theta_{23}$-octant is not affecting the measurement of $\delta$: 
a consequence of the small $\delta$-dependence of the ``octant clone'' location in the ($\theta_{13},\delta$) plane
for $\theta_{13} \geq 1^\circ$, \cite{Donini:2003vz}.

\begin{figure}[t!]
\vspace{-0.5cm}
\begin{center}
\begin{tabular}{cc} 
\hspace{-0.55cm} \epsfxsize7.5cm\epsffile{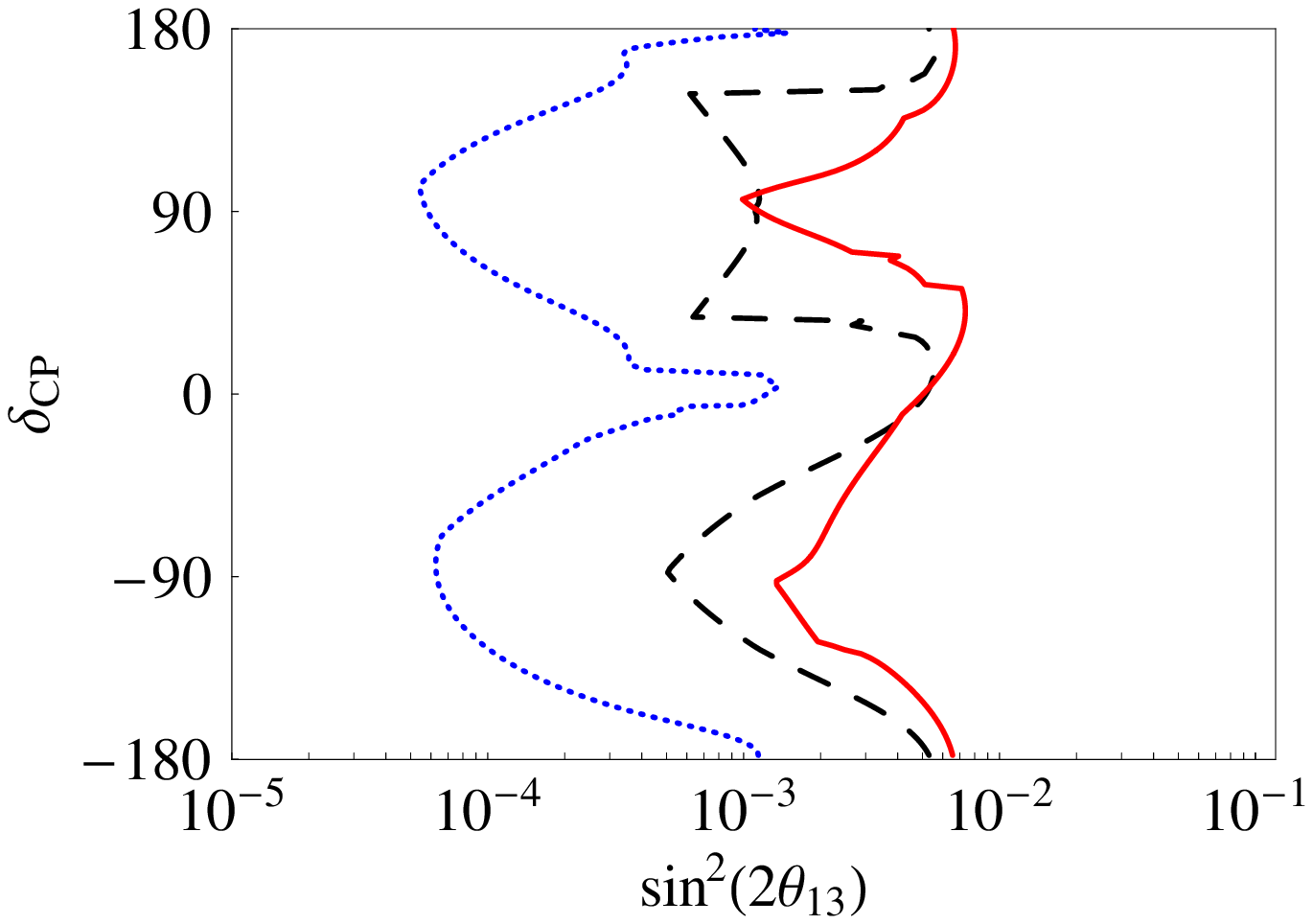} & 
                 \epsfxsize7.5cm\epsffile{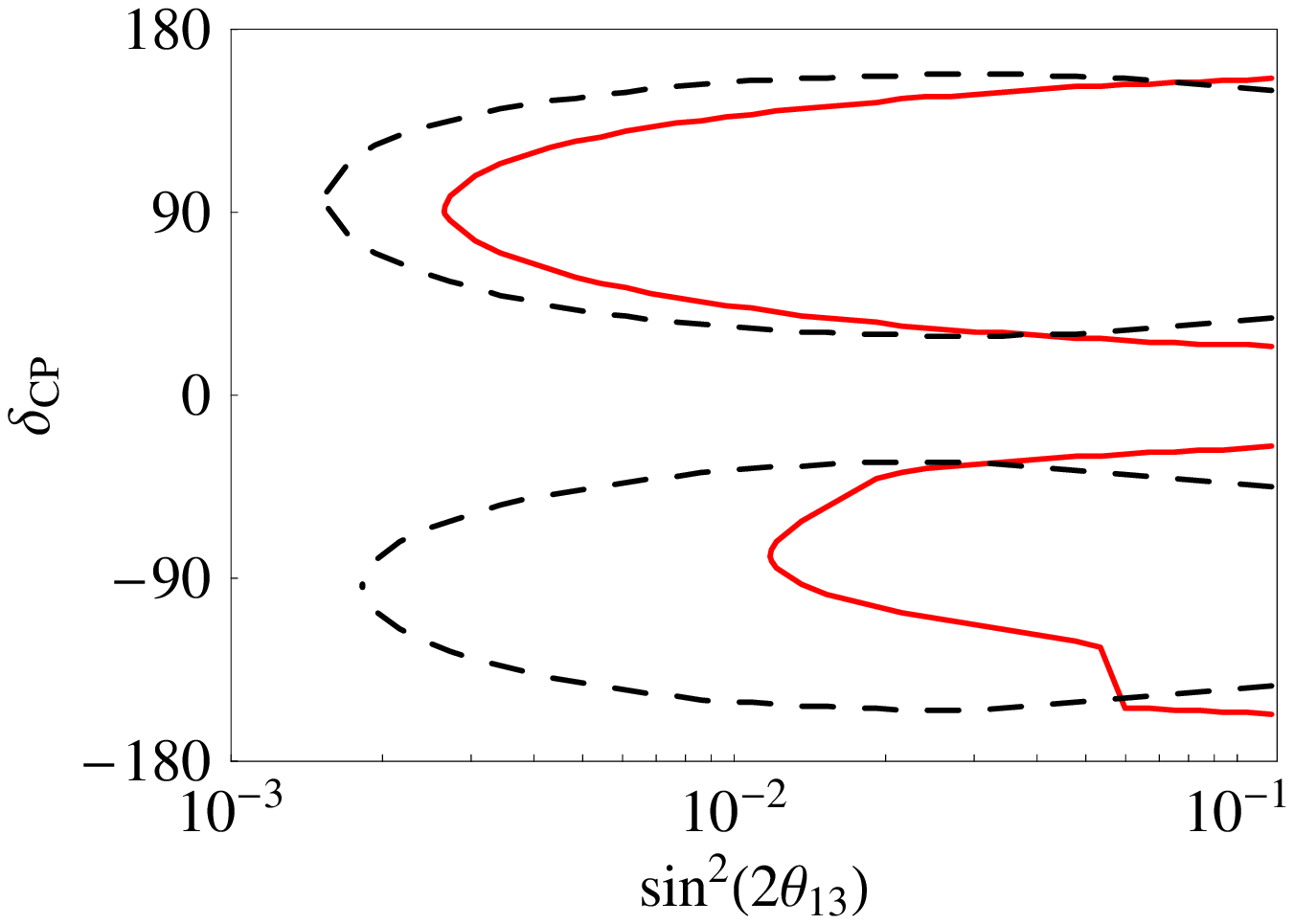}
\end{tabular}
\caption{\it 
Left: 3$\sigma$ $\theta_{13}$-sensitivity for the standard ``low'' $\gamma$ setup (dashed), 
the alternating ions ``low'' $\gamma$ setup (solid) and the standard ``medium'' $\gamma$ setup (dotted); 
Right: CP-discovery potential for the standard ``low'' $\gamma$ setup (dashed) and
the alternating ions ``low'' $\gamma$ setup (solid).
The parametric degeneracies are taken into account as in Ref.~\cite{Donini:2004iv}.
}
\label{fig:sens}
\end{center}
\end{figure}

In Fig.~\ref{fig:sens} we present our results for the $\theta_{13}$-sensitivity (left panel) and 
the CP-discovery potential (right panel), defined as in Refs.~\cite{Donini:2004hu} and \cite{Donini:2004iv}, 
for the standard ``low'' $\gamma$ setup (dashed lines), the alternating ions ``low'' $\gamma$ setup (solid lines) 
and the standard ``medium'' $\gamma$ setup (dotted lines). 
All of the parametric degeneracies are taken into account. Lines represent 3$\sigma$ contours.
As it was expected, our {\it Alternating Ions Scheme} has the worst $\theta_{13}$-sensitivity as a consequence of the lower
statistics (Fig.~\ref{fig:sens}, left). Being the end-point energy for $^8$Li and $^8$B around 13 MeV, 
to tune the corresponding neutrino beams to the first oscillation peak we must increase the baseline 
from $L = 130$ Km to $L = 650$ Km, for fixed $\gamma$. 
Therefore, a smaller statistics is expected in the far detector with respect to the standard ``low'' $\gamma$ setup, 
obtaining thus a reduced $\theta_{13}$-sensitivity. The decrease in statistics due to the longer baseline is 
compensated at the ``medium'' $\gamma$ setup by an increase in $\gamma$. The gain in increasing $\gamma$ is such 
that the $\theta_{13}$-sensitivity of the standard ``medium'' $\gamma$ setup is far better than that of the 
standard ``low'' $\gamma$ one, as it was noticed in Ref.~\cite{Burguet-Castell:2005pa}.
The CP-discovery potential of the {\it Alternating Ions Scheme} is, as expected, worse than that of the standard scenario for low values of 
$\theta_{13}$. In the large $\theta_{13}$ case in which we are interested in, though, our scheme would be able to discover CP-violation  
for a larger fraction of the parameter space. The CP-discovery potential of the ``medium'' $\gamma$ setup is much better 
and it is not shown in this Figure. 

\begin{figure}[t!]
\vspace{-0.5cm}
\begin{center}
\begin{tabular}{cc} 
\hspace{-0.55cm} \epsfxsize7.5cm\epsffile{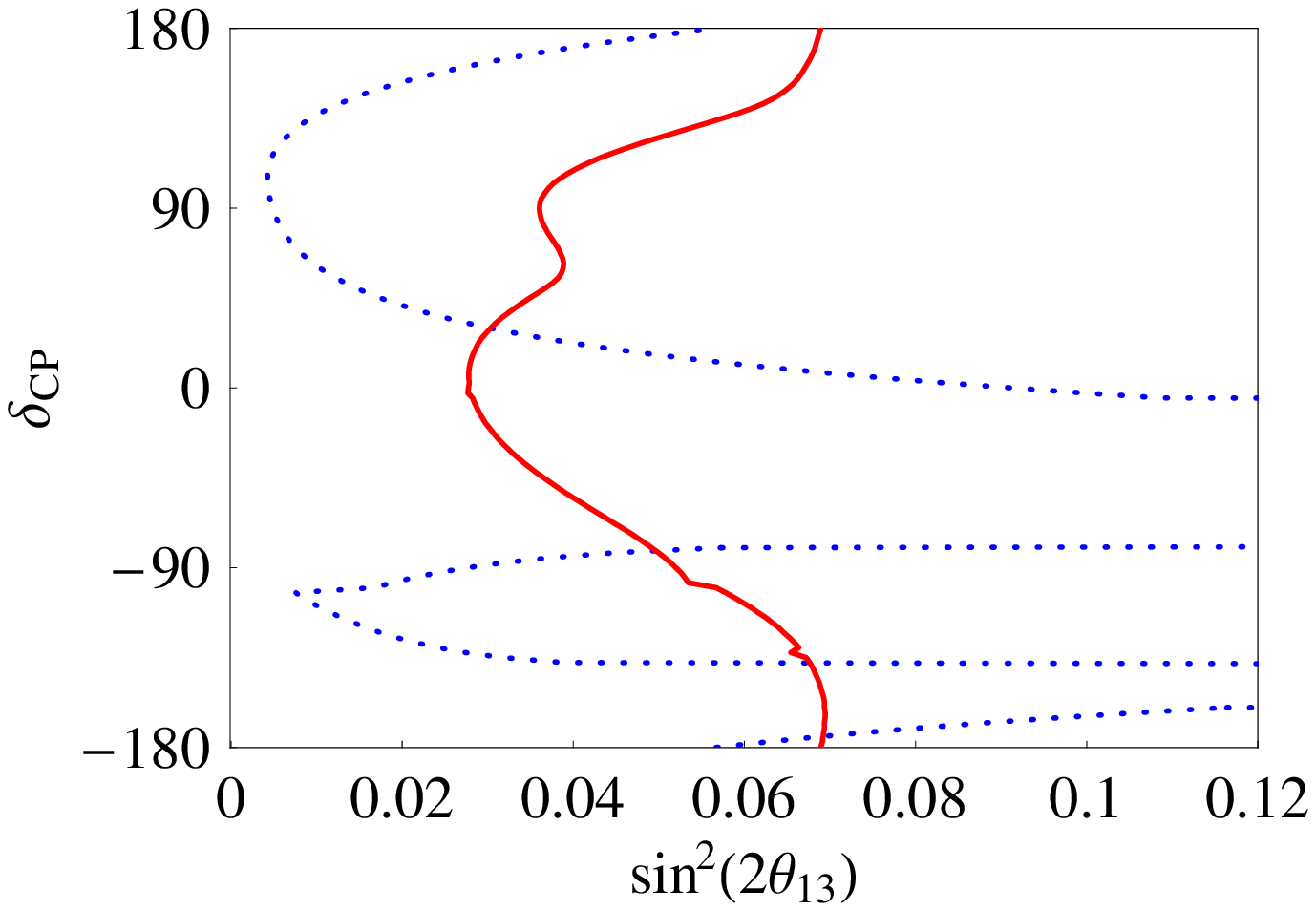} & 
                 \epsfxsize7.5cm\epsffile{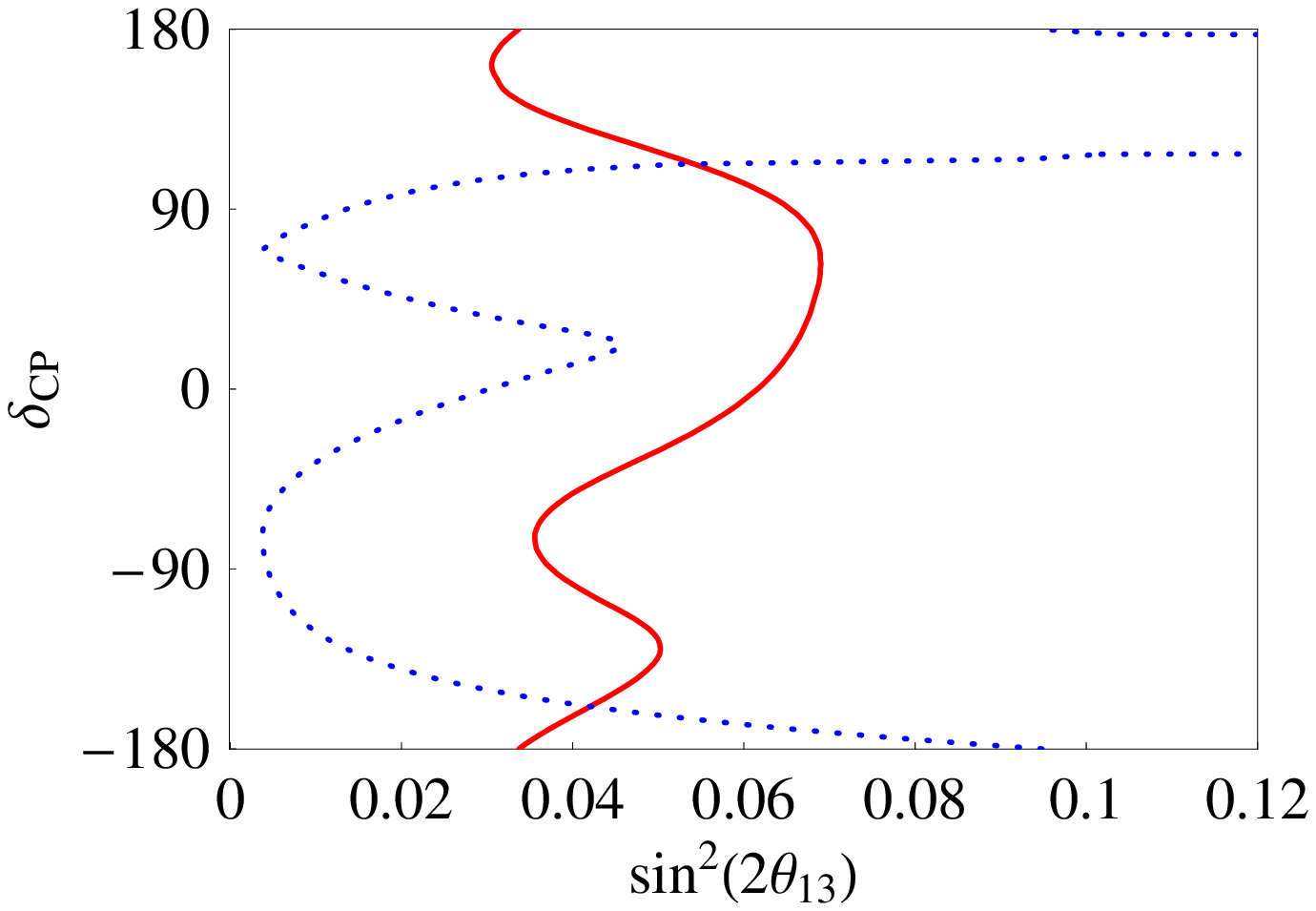}
\end{tabular}
\caption{\it 
Comparison of the sign($\Delta m^2_{23}$)-sensitivity for normal (left panel) and inverted (right panel) hierarchy
for the standard ``low'' $\gamma$ setup (dashed), 
the alternating ions ``low'' $\gamma$ setup (solid) and the standard ``medium'' $\gamma$ setup (dotted).
The parametric degeneracies are taken into account as in Ref.~\cite{Donini:2004iv}.
}
\label{fig:sign}
\end{center}
\end{figure}

In Fig.~\ref{fig:sign} we present our results for the sign($\Delta m^2_{23}$)-sensitivity, for normal (left) and inverted (right) hierarchy,
defined as in Ref.~\cite{Donini:2005db}.
It can be noticed in both cases that the {\it Alternating Ions Scheme} can measure the sign of the atmospheric mass difference 
for $\sin^2 (2 \theta_{13}) \geq 0.04$ ($\theta_{13} \geq 5^\circ$). This must be compared with the standard
``low'' $\gamma$ setup, with no sensitivity to sign($\Delta m^2_{23}$) due to its too short a baseline.
Even compared to the standard ``medium'' $\gamma$ setup, with a much larger statistics, our scheme is particularly
effective for $|\delta| \simeq 0$ ($|\delta| \simeq 180^\circ$) for normal (inverted) hierarchy, a region of the parameter space in which 
other setups are not working very well\footnote{This was the case for the Neutrino Factory, also, see Ref.~\cite{Donini:2005db}.}.
This is a consequence of the combination of neutrino beams tuned at different oscillation peaks, something that
guarantees that ``sign clones'' are located in different regions of the ($\theta_{13},\delta$) parameter space
also in the case of a vanishing $\delta$.

\section{Conclusions}

In this letter we have tried to understand if the combination of neutrino beams originating from the $\beta$-decay
of ions with different end-point energies, such as $^6$He/$^8$Li and $^{18}$Ne/$^8$B, can be used to 
solve the parametric degeneracy in the ($\theta_{13},\delta$) plane. 
The outcome of this analysis is the following: the {\it Alternating Ions Scheme} is
indeed quite effective when the $\gamma$ at which ions can be accelerated is limited to $\gamma \leq 250$, 
e.g. using the present SPS at CERN. In this case, the eightfold-degeneracy is reduced to a twofold one
by solving the so-called ``intrinsic degeneracy'' and by measuring the sign of the atmospheric mass difference
for $\theta_{13} \geq 5^\circ$. This is much better than what is obtained at the {\it standard ``low'' $\gamma$ setup},
with only $^6$He and $^{18}$Ne circulating in the storage ring (although the ultimate sensitivity to $\theta_{13}$ is
somewhat reduced). The main advantage of this scheme is that the neutrino beams originating from the decay of different ions 
have different $L/E$, something than can be effectively used to solve degeneracies. 
The {\it Alternating Ions Scheme} is generally outperformed by setups with larger $\gamma$
(something possible, for example, using a refurbished SPS). However, the {\it standard ``medium'' $\gamma$ setups}
cannot measure the sign of the atmospheric mass difference for small values of $\delta$, something at hand
in our scheme due to the combination of neutrino beams tuned at different oscillation peaks.

In summary, we think that the {\it Alternating Ions Scheme} is particularly well-suited in the physics case
where $\theta_{13}$ is relatively large (i.e. measurable at T2K-I). 

\section*{Acknowledgements}

We thank E.~Couce, M.B.~Gavela, J.J.~G\'omez-Cadenas, P.~Hern\'andez, M.~Lindroos, J.~Men\'endez, P.~Migliozzi, 
C.~Pe\~na-Garay, A.~Poves, S.~Rigolin and O.~Tengblad for the useful discussions.
The authors acknowledge the financial support of MCYT through project FPA2003-04597 and of the European Union
through the networking activity BENE. E.F.M. aknowledges financial support from the Universidad Autonoma de Madrid.

\end{document}